# Radar echo, Doppler Effect and Radar detection in the uniformly accelerated reference frame

Bernhard Rothenstein[1)] and Stefan Popescu[2)]

1) Politehnica University of Timisoara, Physics Department, Timisoara, Romania
2) Siemens AG, Erlangen, Germany

***Abstract.*** *The uniformly accelerated reference frame described by Hamilton[7] and Desloge and Philpott[8] involves the observers $R_i(g_i)$ who perform the hyperbolic motion with constant proper acceleration $g_i$. They start to move from different distances $c^2/g_i$ measured from the origin O of the inertial reference frame K(XOY), along its OX axis with zero initial velocity. Equipped with clocks and light sources they are engaged with each other in Radar echo, Doppler Effect and Radar detection experiments. They are also engaged in the same experiments with an inertial observer $R_0(0,0)$ at rest in K(XOY) and located at its origin O. We derive formulas that account for the experiments mentioned above. We study also the landing conditions of the accelerating observers on a uniformly moving platform.*

### 1. Introduction

A ticking clock displays a given time that represents its reading. Identical clocks located at the different points of an inertial reference frame, synchronised in accordance with a synchronization procedure proposed by Einstein[1] display the same running time. Consider the clock $C_1(x_1,0)$ located at the point $M_1(x_1,0)$ of the OX axis of the K(XOY) inertial reference frame and an event that takes place in front of it when it reads $t_1$. We say that an event $E_1(x_1,0,t_1)$ took place, characterized by its space coordinates $(x_1,0)$ and by a time coordinate $t_1$. An event $E_1(x_1,0,t_2)$ takes place in front of the same clock when it reads $t_2$. We say that the two events are separated by a time interval $\Delta t = t_2 - t_1$, but take place at the same point in space. An important part is played by the clock $C_0(0,0)$ located at the origin O of K.

Special relativity becomes involved when we consider a second inertial reference frame K'(X'O'Y'). The axes of the two frames are parallel to each other, the OX(O'X') axes are overlapped and K' moves relative to K with constant speed **V** in the positive direction of the OX(O'X') axes. Along the O'X' axis we find the clocks **C'(x',0)** synchronized in accordance with the same synchronization procedure as in K, such that all of them display the same running time **t'**. When the origins O and O' of K and K', respectively, are located at the same point in space, all the clocks involved read **t=t'=0.** The notation **E' (x',0,t')** stands for an event that takes place at the point **M'(x',0)** when clock **C'(x',0)** located there reads **t'**. Events $E'_1(x',0,t'_1)$ and $E'_2(x',0,t'_2)$ take place at the same point in space, but are separated by a time interval $\Delta t' = t'_2 - t'_1$.

In accordance with the second postulate of the special relativity, the light signals that perform the synchronization of the clocks in the two reference frames propagate with the same speed **c** in all inertial reference frames.

The scenario we follow requires that at a given point of the overlapped axes OX(O'X') we find a clock **C(x,0)** of the K frame reading **t** and a clock **C'(x',0)** of the K' frame reading **t'**. A fundamental problem in special relativity theory is to find a relationship between **x** and **x'** and between **t** and **t'**. This problem is associated with the following relativistic effects: **radar echo, time dilation** and **Doppler Effect.**[2]

## 2. The radar echo

### 2.1 The radar echo with uniform moving targets
The **radar echo experiment** [2] involves a single clock of a single inertial reference frame, say K. The simple experiment we propose involves a source of light **S** located at the origin O of frame K and in a state of rest relative to it. A mirror **M** moves with constant velocity **V** along the OX axis. It is located in front of the source at a time **t=0**. A light signal emitted at a time $t_e$ arrives at the location of the moving mirror at a time $t_m$ being instantly reflected back by the mirror and returning to the source at a time $t_r$. The events involved in the experiment are displayed by the classical space-time diagram[4] presented in **Figure 1** in the case of an approaching mirror $M_a$ and respectively of a receding mirror $M_r$.

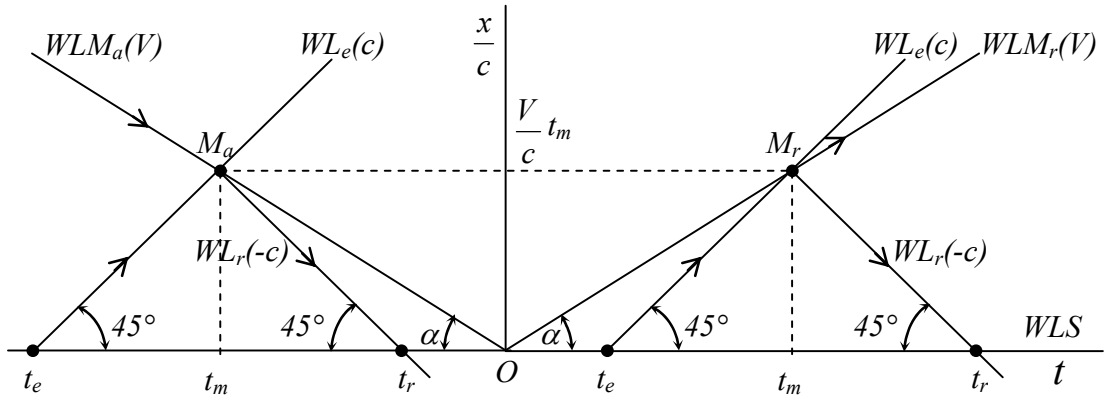

**Figure 1.** The radar echo experiment displayed by the space-time diagram. Approaching mirror - left, receding mirror - right.

The world lines of the light signals emitted by the source $WL_e$ make an angle of $45^0$ with the positive direction of the time axis. The world lines of the light signals reflected on the mirror $WL_r$ make an angle of $-45^0$ with the positive direction of the time axis. The world line of the receding mirror $WLM_r$ makes an angle $\alpha$ ($\tan\alpha = V/c$), whereas the world line of the approaching mirror $WLM_a$ makes an angle $-\alpha$ with the positive direction of the time axis. We calculate the reception time as function of emission time in the case of the receding mirror. Simple geometry applied to Figure 1 leads to :

$$t_r = t_m + \frac{V}{c}t_m \qquad (1)$$

$$t_e = t_m - \frac{V}{c} t_m \tag{2}$$

from where, eliminating $t_m$, we obtain:

$$t_r = \frac{1+V/c}{1-V/c} \cdot t_e \tag{3}$$

In the case of the approaching mirror in the same way we obtain:

$$t_r = \frac{1-V/c}{1+V/c} \cdot t_e \tag{4}$$

Here $V$ represents the unsigned, absolute value of the speed of the mirror. Consider now that the light source **S** emits periodic signals having the constant period $T_e = \Delta t_e$. These signals will be reflected on the mirror at periodic time intervals $T_m = \Delta t_m$ and arrive back at source at periodic time intervals $T_r = \Delta t_r$.

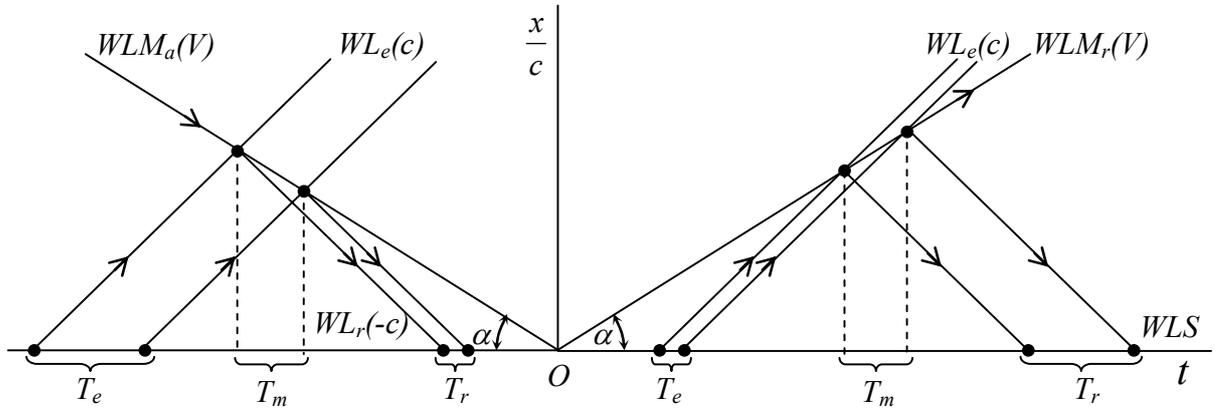

**Figure 2.** Illustration of the frequency shift occurring in the radar echo experiment. Approaching mirror - left, receding mirror - right.

We illustrate in **Figure 2** the well known variation of reception period as function of mirror speed **V** and its moving direction relative to the source. The reception period may be obtained by differentiating the equations (3) and (4). For a mirror receding the source at constant speed we get the following relation:

$$\frac{T_r}{T_e} = \frac{1+V/c}{1-V/c}. \tag{3a}$$

In the case of the approaching mirror in the same way we obtain

$$\frac{T_r}{T_e} = \frac{1-V/c}{1+V/c}. \tag{4a}$$

The experiment described above involves a single observer $R_0(0,0)$ and his wrist watch $C_0(0,0)$. As we can see, the formulas that describe the radar echo experiment are time independent, due to the fact that the velocity of the mirror relative to the observer is constant.

### 2.2 The radar echo with accelerated targets

The problem is to find what happens when the mirror performs the so called hyperbolic (constant proper acceleration) motion[7] which best mimics the scenario presented in **Figure 1**. In this case the axial position of the mirror is described by the following equation:

$$x = \frac{c^2}{g}\sqrt{1+\frac{g^2 t^2}{c^2}} = \frac{c^2}{g}\cosh\frac{g}{c}t' \qquad (5)$$

where **g** represents the constant proper acceleration of the mirror, **t** represents the readings of the synchronised clocks of the K frame when the mirror is located successively in front of them and **t'** represents the reading of the wrist watch of an observer **R'** commoving with the mirror. The velocity of the mirror is given by :

$$V = \frac{gt}{\sqrt{1+\frac{g^2 t^2}{c^2}}} = c\tanh\frac{g}{c}t' \qquad (6)$$

Here **V** is the signed value of the mirror speed. For the approaching mirror we have **t** < 0, **t'** < 0 and consequently **V** < 0, whereas for the receiving mirror we have respectively **t** > 0, **t'** > 0 and **V** > 0.

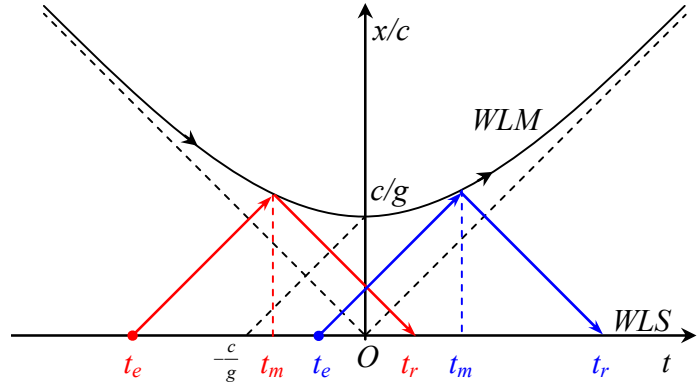

**Figure 3.** The radar echo experiment with accelerating mirrors.
Approaching mirror - left, receding mirror - right.

**Figure 3** shows the world line of the mirror *WLM* and the world lines of the successive light signals emitted by the stationary source and reflected by the moving mirror. The mirror comes decelerating from **x=+∞** where it is located at **t=t'=-∞** and where its velocity was **–c,** and it arrives with zero velocity at the point **$x_0=c^2/g$** at a time **t=t'=0**. The mirror then accelerates towards **x=+∞**, where it arrives at a time **t=t'=∞** with speed **+c**. As we can see, the light signals emitted in the time interval $-\infty < t_e < -c/g$ are reflected by an approaching mirror, whereas the light signals emitted in the time interval **–c/g < $t_e$ < 0** are reflected by an receding mirror. Only the light signals emitted at negative times **$t_e \leq 0$** get the chance to reach the mirror and generate a reception event at always positive reception times **$t_r \geq 0$**.

In a first approach to solve the problem we use the Cochran[3] hypothesis, which is valid when the time interval between the receptions of two successive light signals is small enough such that during their reception the speed of the observer remains approximately constant. Under such conditions and because the incidence and the reflection of the light signal take place at the same point in space and at a given time, we obtain the relationship between the emission and reception times and respectively periods by replacing in (3) and (3a) the **instantaneous value of V** given by (6). By this way these expression became time dependent and we can express their time dependence as a function of **t'**, which is the time displayed by the clock commoving with the mirror when the reflection takes place on it, or as a function of the time **t**, which is the time displayed by the clocks of the rest frame of the source when the same event takes place. In the first case we replace the right side of **V** expression (6) in (3a) and obtain:

$$\frac{t_r}{t_e} = \frac{T_r}{T_e} = \exp\left(2\frac{g}{c}t'\right) \tag{7}$$

where **t'** changes in the range of values $-\infty < t' < +\infty$.

Expressing **V** by the left side of (6), which is a function of **t**, we obtain:

$$\frac{t_r}{t_e} = \frac{T_r}{T_e} = \frac{1 + c^{-1}gt/\sqrt{1 + g^2t^2/c^2}}{1 - c^{-1}gt/\sqrt{1 + g^2t^2/c^2}}, \tag{8}$$

where **t** changes in the range $-\infty < t < 0$.

In a second but more accurate and general valid approach we use the equation of movement (5) to calculate the time $t_m$ when the light signal reaches the mirror. Starting with the expressions for the position $x_m$ where the reflection takes place on mirror:

$$x_m = c(t_m - t_e) = \frac{c^2}{g}\sqrt{1 + \frac{g^2 t_m^2}{c^2}} \tag{9}$$

we get the expression for the reflection time $t_m$ as:

$$t_m = \frac{t_e^2 - c^2/g^2}{2t_e}. \tag{10}$$

Furthermore from the obvious relation $2t_m = t_e + t_r$ we get the expression of $t_r$ as:

$$t_r = \frac{-c^2/g^2}{t_e}. \tag{11}$$

As we can see from **Figure 3**, the last signal that participates in the radar echo experiment is emitted at a time $t_e$=**0**. Equation (11) reveals the important fact that the product of emission and reception times $t_e \cdot t_r$ is time independent, depending only on the constant proper acceleration **g** of the mirror.

If the source **S** emits periodic light signals having the constant period $T_e$ then the signals reflected on the mirror will arrive back at source at periodic time intervals $T_r$. By differentiating (11) we obtain the relation between the emission and reception periods as:

$$\frac{T_r}{T_e} = \frac{c^2/g^2}{t_e^2}. \tag{12}$$

Equation (11a) reveals the interesting fact that the reception period continuously increases (red shift) from $T_r = 0$ at $t_e = -\infty$ occurring with an approaching source up to $T_r = \infty$ for $t_e = 0$ with a receding source when the radar echo vanishes. Furthermore we distinguish the following relations between the emission and reception periods:
- $T_r < T_e$ when $t_e < -c/g$ and the reflection takes place on an approaching mirror
- $T_r = T_e$ when $t_e = -c/g$ and the reflection takes place on a stationary mirror
- $T_r > T_e$ when $t_e > -c/g$ and the reflection takes place on a receding mirror.

### 3. The Doppler Effect

**3.1. The Doppler Effect with uniform moving observers. Locality and non-locality in the reception of periodic light signals**[4,5,6]

Consider a source of light **S(0,0)** located at the origin O of its rest frame K and the clocks **C(x,0)** located at the different points of the OX axis, synchronized with clock $C_0(0,0)$ in accordance with Einstein's clock synchronization procedure. When all of the clocks mentioned above read $t_e$, source **S(0,0)** emits a light signal in the positive direction of the OX axis which arrives at the location of the clock **C(x,0)** when all the clocks read $t_r$; therefore, it is obvious that

$$t_r = t_e + x/c. \tag{13}$$

We allow a **very small** change in the readings of the clocks involved, which reveals that they are related by

$$dt_r = dt_e + \frac{dx}{c}. \tag{14}$$

Taking into account that by definition

$$V = \frac{dx}{dt_r} \tag{15}$$

represents the instantaneous velocity measured at the reception time of an observer **R'** attached to the clock **C(x,0)** then (14) leads to

$$\frac{dt_e}{dt_r} = 1 - \frac{V}{c}. \tag{16}$$

A clock **C'(x',0)** attached to the moving observer **R'** registers a change in the reading of his clock $dt'_r$ related to $dt_r$ by the time dilation formula

$$dt_r = \frac{dt'_r}{\sqrt{1 - \frac{V^2}{c^2}}} \tag{17}$$

with which (15) becomes

$$\frac{dt_e}{dt'_r} = \sqrt{\frac{1 - V/c}{1 + V/c}} \tag{18}$$

We consider now that the source **S(0,0)** emits successive light signals at constant proper time intervals $dt_e$. In accordance with (18), observer **R'** receives them at constant proper time intervals $dt'_r$. Equation (18) holds exactly only in the case of **very small periods (very high frequencies)** for which the instantaneous velocity of **R'** does not change significantly between the reception of two successive light signals and the observer **receives two successive light signals from the same point in space.** Relativists describe these particular circumstances by saying that **locality takes place in the time interval measurement** [4,5]. If the observer is moving with constant velocity then (17) holds in the case of finite emission and reception periods of successive light signals i.e.

$$\frac{\Delta t_e}{\Delta t'_r} = \frac{T_e}{T'_r} = \sqrt{\frac{1-V/c}{1+V/c}} \ . \tag{19}$$

Here $T_e = \Delta t_e$ is the constant proper emission period as measured in source frame and $T_r' = \Delta t_r'$ is the proper reception period as measured by the moving observer. We discover in this case the time independence of the Doppler shift relation.

### 3.2. The Doppler Effect with accelerating observers

But what happens if the motion of the clock attached to the observer is not uniform? Cochran[6] considers that the time interval between the receptions of two successive light signals is small enough to consider that during their reception the speed of the observer remains constant, such that the two successive light signals are received from the same point in space. Under such conditions Cochran[6] simply replaces **V** in (23) with its instantaneous value (6). We obtain the corresponding results by observing that $T_e$ and $T_r$ in the radar echo experiment and $T_e$ and $T_r'$ in the Doppler effect are related by

$$D = \frac{T_e}{T'_r} = \sqrt{\frac{T_r}{T_e}} = \exp\frac{gt'}{c} = \sqrt{\frac{1+\frac{gt}{c\sqrt{1+\frac{g^2t^2}{c^2}}}}{1-\frac{gt}{c\sqrt{1+\frac{g^2t^2}{c^2}}}}} \ . \tag{20}$$

where equation (24) holds only in the case of the locality assumption, when we consider that two successive light signals are received by the moving observer from the same point in space (locality assumption) a condition that is favoured by "very small periods" and "low velocities".

The Doppler Effect is characterized by the Doppler factor defined as

$$D = \frac{f'_r}{f_e} \tag{21}$$

where $f_e = 1/T_e$ and $f_r' = 1/T_r'$ represent the frequencies at which the source emits successive light signals and the observer receives them, respectively. This justifies the notation in (20).

In a more realistic and accurate approach, we parameterise the experiment by considering that the source emits at regular time intervals $NT_e$ **(N<0).** The position of the observer when he receives the $N^{th}$ emitted light signal is defined by

$$x = \frac{c^2}{g}\cosh\frac{g}{c}t'_{r,N} \tag{22}$$

whereas the position of the $N^{th}$ emitted light signal is defined by

$$x = c\left(\frac{c}{g}\sinh\frac{g}{c}t'_{r,N} - NT_e\right). \tag{23}$$

Eliminating **x** between (22) and (23), we obtain

$$t'_{r,N} = \frac{c}{g}\ln\left(-\frac{g}{c}NT_e\right) \tag{24}$$

an equation with physical meaning because, as we have mentioned above, only the light signals emitted in the time interval $-\infty < NT_e < 0$ take part in the Doppler Effect experiment. The time interval between the receptions of two successive light signals is given by

$$T'_r = \Delta t'_{r,N-1,N} = t'_{r,N} - t'_{r,N-1} = \frac{c}{g}\ln\frac{N-1}{N} \tag{25}$$

and we underline two of its interesting properties:
- **it does not depend on the period $T_e$ at which the source emits the successive light signals.**
- **it shows a continuous variation of $T'_r$**

We present in **Figure 4** the variation of $T'_r$ with **N** for different values of T=**c/g.**

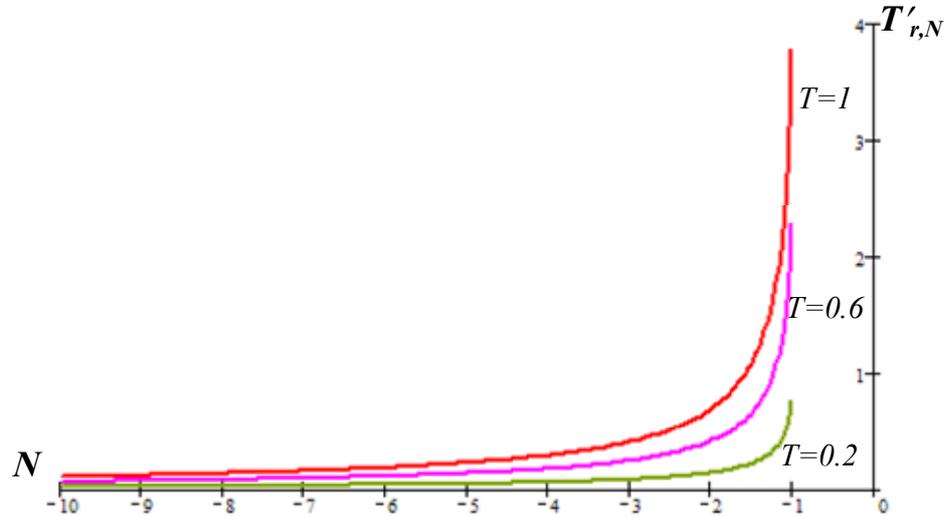

**Figure 4.** The variation of the time interval $T'_r$ between the reception of two successive light signals by the accelerating receiver vs. the order number **N** of the light signal.

**4. Radar echo and Doppler shift in the uniformly accelerating reference frame**

The observers $R'_1(g_1)$ and $R'_2(g_2)$ moving with the constant proper accelerations $g_1$ and $g_2$ are the components of the **uniformly accelerating reference frame** if at **t=0** both are at rest at the distances $c^2/g_1$ and $c^2/g_2$ from the origin O of the stationary reference frame and start to perform the hyperbolic motion

$$x = \frac{c^2}{g}\cosh\frac{g}{c}t' \qquad (29)$$

Some of the properties of the **uniformly accelerating reference frame** are presented by Hamilton[7] and by Desloge and Philpott[8]. An important property of this reference frame consists of the fact that the distance between two observers is time independent. The observer who starts his trip from the point **x=0** moves with infinite acceleration.

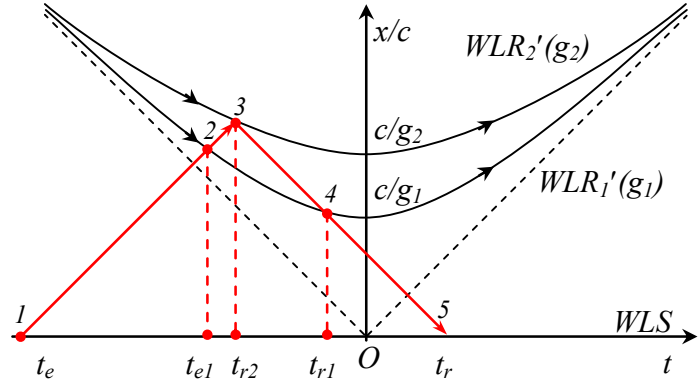

**Figure 5.** The radar experiment with two accelerating observers. Approaching mirror - left, receding mirror - right.

We present the world lines of the accelerating observers in **Figure 5**. Consider that the two observers mentioned above are each involved in a Doppler Effect experiment with the stationary observer $R_0$. $R_0$ emits two successive light signals, labelled **–(N-1)** and **–N**, and separated by a time interval $T_e$. $R'_1(g_1)$ receives them separated by a time interval

$$T'_{r,N-1,N,g_1} = \frac{c}{g_1}\ln\frac{N-1}{N} \qquad (30)$$

Let consider now that $R'_1(g_1)$ emits further the received signals without delay, therefore separated by the emission intervals $T'_{e,N-1,N,g_1} = T'_{r,N-1,N,g_1}$. Then $R'_2(g_2)$ receives the same light signals separated by a time interval:

$$T'_{r,N-1,N,g_2} = \frac{c}{g_2}\ln\frac{N-1}{N}. \qquad (31)$$

If we characterise the Doppler Effect that involves the two observers of the same uniformly accelerating reference frame then the Doppler factor will be:

$$D = \frac{T'_{e,N-1,N,g_1}}{T'_{r,N-1,N,g_2}} = \frac{g_2}{g_1} < 1 \qquad (32)$$

and we observe that a red shift takes place.

As we can see, the Doppler factor D is time independent in this case, mainly because the distance between the observers involved here is time independent as well. We have not found until now this result mentioned in the literature on the subject.

Let suppose now that observers $R'_1$ and $R'_2$ are engaged in radar echo experiment illustrated in **Figure 6**. In this case observer $R'_1$ is performing the active role in the radar experiment.

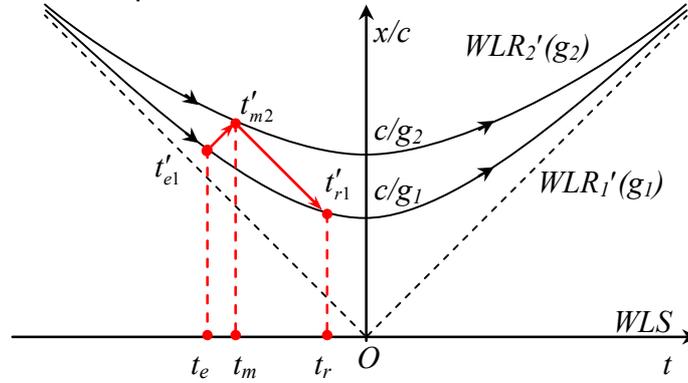

**Figure 6.** The radar experiment with two accelerating observers. Observer $R'_1$ performs the experiment.

The following events are involved here:
- $E_e(x_e,t_e)$ → Observer $R'_1$ emits a light signal at position $x_e$ and time $t_e$ (proper time $t'_{e1}$) towards observer $R'_2$;
- $E_m(x_m,t_m)$ → Observer $R'_2$ receives the light signal at position $x_m$ and time $t_m$ (proper time $t'_{m2}$) and reflects it back without delay towards observer $R'_1$;
- $E_r(x_r,t_r)$ → Observer $R'_1$ receives the reflected light signal at position $x_r$ and time $t_r$ (proper time $t'_{r1}$) and calculates the radar detected coordinates of $R'_2$.

In the inertial frame at rest we have the following obvious relations:
$$\begin{aligned} c(t_m - t_e) &= x_m - x_e \\ c(t_r - t_m) &= x_m - x_r \end{aligned} \qquad (33)$$

which rearranges for the respective events become
$$\begin{aligned} x_e - ct_e &= x_m - ct_m \\ x_r + ct_r &= x_m + ct_m \end{aligned} \qquad (34)$$

Transposing in proper time coordinates of the respective observers:

$$\frac{1}{g_1}\left[\cosh\left(\frac{g_1}{c}t'_{e1}\right)-\sinh\left(\frac{g_1}{c}t'_{e1}\right)\right]=\frac{1}{g_2}\left[\cosh\left(\frac{g_2}{c}t'_{m2}\right)-\sinh\left(\frac{g_2}{c}t'_{m2}\right)\right]$$

$$\frac{1}{g_1}\left[\cosh\left(\frac{g_1}{c}t'_{r1}\right)+\sinh\left(\frac{g_1}{c}t'_{r1}\right)\right]=\frac{1}{g_2}\left[\cosh\left(\frac{g_2}{c}t'_{m2}\right)+\sinh\left(\frac{g_2}{c}t'_{m2}\right)\right]$$

(35)

and expanding the hyperbolic terms we get the simplified relations:

$$\frac{1}{g_1}\exp\left(-\frac{g_1}{c}t'_{e1}\right)=\frac{1}{g_2}\exp\left(-\frac{g_2}{c}t'_{m2}\right)$$

$$\frac{1}{g_1}\exp\left(\frac{g_1}{c}t'_{r1}\right)=\frac{1}{g_2}\exp\left(\frac{g_2}{c}t'_{m2}\right)$$

(36)

By definition, observer $R'_1$ assigns to the reflection event that takes place on the world line of observer $R'_2$ a radar detected distance to target:

$$X'_{1,radar}=\frac{c}{2}(t'_{r,1}-t'_{e,1})$$

(37)

and a radar detected target time coordinate:

$$t'_{m1,radar}=\frac{1}{2}(t'_{r,1}+t'_{e,1})$$

(38)

Multiplying the terms of (36) we obtain

$$X'_{1,radar}=\frac{c^2}{g_1}\ln\frac{g_1}{g_2}$$

(39)

resulting that the distance in-between observers as detected by $R'_1$ is time independent. Furthermore this distance is positive because $g_1>g_2$ by the convention we made in **Figure 6**. If $g_1 = g_2$ then the distance occurs to be always zero as the two accelerating observes permanently superpose their positions.

The distance between observers $R'_2$ and $R'_1$ measured in the uniformly accelerated reference frame as a difference between their proper space coordinates is

$$X'=\frac{c^2}{g_2}-\frac{c^2}{g_1}$$

(40)

Equation (39) compared to (40) reveal an important property of the uniformly accelerating reference frame, namely that distances measured using different detection procedures are not equal. Desloge and Philpott[8] also obtained this result.

Dividing the terms of (36) we obtain

$$t'_{m1,radar} = \frac{g_2}{g_1} t'_{m2} \tag{41}$$

the relationship between the two times being linear. Furthermore we notice that because $g_1 > g_2$ we have a time compression effect i.e. the "radar clock" of the faster moving observer $R'_1$ is running slower. If $g_1 = g_2$ then the time readings coincide as the two accelerating observes are always situated at the same point in space.

Lets now reverse the scenario above and suppose that observers $R'_2$ is performing the active role in the radar detection experiment. This new experiment is illustrated by figure below.

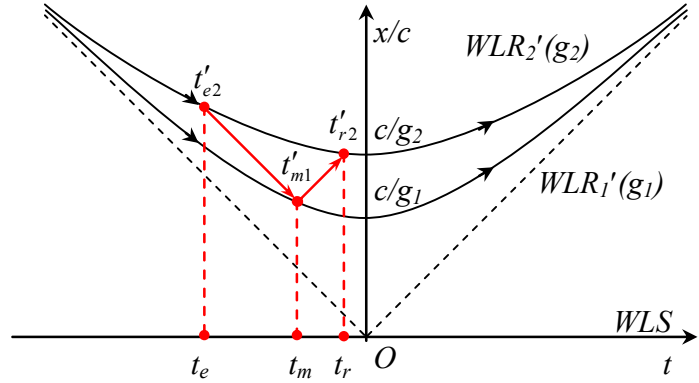

**Figure 7.** The radar experiment with two accelerating observers. Observer $R'_2$ performs the experiment.

In the inertial frame at rest we have the following obvious relations:
$$\begin{aligned} c(t_m - t_e) &= x_e - x_m \\ c(t_r - t_m) &= x_r - x_m \end{aligned} \tag{42}$$

Following the same calculation method as above we get the relations:
$$\frac{1}{g_2}\exp\left(\frac{g_2}{c}t'_{e2}\right) = \frac{1}{g_1}\exp\left(\frac{g_1}{c}t'_{m1}\right)$$
$$\frac{1}{g_2}\exp\left(-\frac{g_2}{c}t'_{r2}\right) = \frac{1}{g_1}\exp\left(-\frac{g_1}{c}t'_{m1}\right) \tag{43}$$

Again by definition, observer $R'_2$ assigns to the reflection event that takes place on the world line of observer $R'_1$ a radar detected distance to target:
$$X'_{2,radar} = \frac{c}{2}\left(t'_{r,2} - t'_{e,2}\right) \tag{44}$$

and a radar detected target time coordinate:
$$t'_{m2,radar} = \frac{1}{2}\left(t'_{r,2} + t'_{e,2}\right) \tag{45}$$

Multiplying the terms of (43) we obtain
$$X'_{2,radar} = \frac{c^2}{g_2} \ln \frac{g_1}{g_2} \tag{46}$$
resulting that the distance in-between observers as detected by $R'_2$ is positive and time independent.

Dividing the terms of (43) we obtain
$$t'_{m2,radar} = \frac{g_1}{g_2} t'_{m1} \tag{47}$$
Again because $g_1 > g_2$ we have a time dilation effect i.e. the "radar clock" of the slower moving observer $R'_2$ is running faster.

Comparing equations (39) and (46) we reveal a new important property of the uniformly accelerating reference frame, namely that distances measured using the radar detection method are different depending on which observer actually performs the radar detection. Dividing these equations we get the result:
$$\frac{X'_{1,radar}}{X'_{2,radar}} = \frac{g_2}{g_1} \tag{48}$$
Because $g_1 > g_2$ we notice a "distance compression" effect i.e. the faster moving observer detects shorter distances by radar detection.

From equations (41) and (47), which gives the relation between the proper time of the same event (the light reflection) for the two accelerating receivers, we get a common form for time transformation as:
$$t'_{m1,radar} \cdot g_1 = t'_{m2,radar} \cdot g_2 \tag{49}$$
whereas for the distances we have a similar from:
$$X'_{1,radar} \cdot g_1 = X'_{2,radar} \cdot g_2 \tag{50}$$

Let consider now the reflection event $E_m(x_m, t_m)$ as seen from the stationary frame. By writing the expression for its coordinates $(x_m, t_m)$ as function of the proper time of the accelerating observers involved we get:
$$\frac{1}{g_1}\cosh\left(\frac{g_1}{c}t'_{m1}\right) = \frac{1}{g_2}\cosh\left(\frac{g_2}{c}t'_{m2}\right)$$
$$\frac{1}{g_1}\sinh\left(\frac{g_1}{c}t'_{m1}\right) = \frac{1}{g_2}\sinh\left(\frac{g_2}{c}t'_{m2}\right) \tag{51}$$
or further:
$$t'_{m1}g_1 = t'_{m2}g_2 + c \ln \frac{g_1}{g_2} \tag{52}$$
which is again a result different than the one obtained by the radar experiment.

## 5. Landing the observers of the uniformly accelerated reference frame

We devote this chapter to the problem of the smooth landing of the observers $R'_i(g_i)$ who make up the uniformly accelerated reference frame. Consider an endless platform that moves with constant velocity $V_0$ in the positive direction of the OX axis of the K inertial reference frame. The origin of the inertial reference frame coincides with the centre of the platform. The observer mentioned above reaches the velocity $V_0$ at a time $t_i$ and at that very time he can land smoothly on the platform defined above. Doing so the reference frame K'(X'O'Y') attached to the platform becomes his rest frame. It is obvious that the different observers will land on the platform at different points and at different times. Once landed on the platform, the clocks $C'_i$ of the different observers become inertial ones displaying time in the way in which such clocks do.

We determine first the time $t_i$ at which the velocity of the considered observer reaches the magnitude $V_0$. From

$$V_0 = \frac{g_i t_i}{\sqrt{1 + \frac{g_i^2 t_i^2}{c^2}}} \tag{53}$$

we obtain

$$t_i = \frac{V_0}{g_i} \frac{1}{\sqrt{1 - V_0^2/c^2}} \tag{54}$$

the landing of the observer taking place at a point, characterized in the K frame, by a space coordinate

$$x_i = \frac{c^2}{g_i} \frac{1}{\sqrt{1 - V_0^2/c^2}} \quad . \tag{55}$$

The Lorentz-Einstein transformations for the space-time coordinates of the same event detected by observers from K and K' respectively enable us to find out a relationship between the space-time coordinates detected from K($x_i$,$t_i$) and detected from K'($x'_i, t'_i$) respectively

$$x'_i = \frac{c^2}{g_i} \tag{56}$$

$$t' = 0. \tag{57}$$

Equations (49) and (50) reveal an interesting property of the uniformly accelerating reference frame: **In all inertial reference frames in which we transport smoothly the observers that make up the uniformly accelerating reference frame, the distance between them remains constant and the clocks remain synchronized.**

With the notations $\beta = V_0/c$ the coordinates of the landing event ($x_i$,$t_i$) become:

$$\frac{x_i}{c} = \frac{c}{g_i} \frac{1}{\sqrt{1-\beta^2}} \quad \text{and} \quad t_i = \frac{c}{g_i} \frac{\beta}{\sqrt{1-\beta^2}} \tag{58}$$

and we represent in a world diagram the landing events as a function of **g** for different values of β.

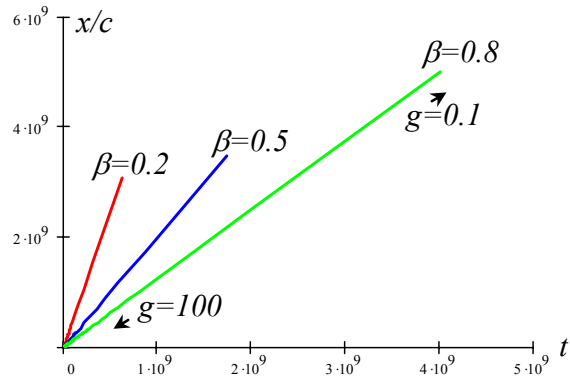

**Figure 8.** The world coordinates of the landing event.

## 6. Conclusions

Extending the results obtained by Hamilton[7] and Desloge and Philpott[8] we present some new results concerning the radar echo, Doppler Effect and Radar detection in the uniformly accelerating reference frame. The landing conditions of the observers who make up the reference frame are also studied.